\documentstyle[12pt]{article}
\newcommand{\dis}{\displaystyle}

\begin{document}
\begin{flushright}
CU/TP-97-3\\
April 1997
\end{flushright}
\vspace{1cm}
\centerline{\LARGE{ Neutrino Mass and Mixing in the Universal }}
\vspace{0.5cm}
\centerline{\LARGE{ Yukawa Coupling Framework }}
\vspace{1cm}
\centerline{\large{ Tadayuki {\sc Teshima}\footnote{E-mail address: teshima@isc.chubu.ac.jp} , Toyokazu {\sc Sakai} and Osamu {\sc Inagaki}}}
\vspace{1cm}
\centerline{\large{\it  Department of Applied Physics, Chubu University}}
\vspace{0.5cm}
\centerline{\large{\it Kasugai, Aichi 487, Japan}}
\vspace{1cm}
\centerline{April 1997} 
\vspace{2cm}
\setlength{\baselineskip}{0.635cm}
\centerline{ABSTRACT}
\par
We present an analysis of neutrino oscillations which are consistent with current solar, atmospheric and terrestrial experiments in a framework with three neutrino mixing. The solutions satisfying these experiments have the large mixing (not maximal mixing) between $\nu_e$-$\nu_{\mu}$ or $ \nu_{\mu}$-$\nu_{\tau}$ in $m_3 \sim {\rm several\ eV}$, where $m_3$ is the heaviest neutrino mass. From these results, we conjecture a form for the mass matrices of charged lepton and neutrino which are described by the universal Yukawa coupling with small violations.

\newpage
\baselineskip=0.635cm
\section{Introduction}
\par
The problem of neutrino masses and oscillations is one of the most interesting issues to study physics beyond the standard model (SM).\cite{STANDARD} In many experiments which are under way, indications in favor of neutrino masses and oscillations have been obtained. In these experiments, the solar neutrino experiments \cite{SOLAR} measure the event rates significantly lower than the ones predicted by the standard solar model, and the atmospheric neutrino experiments \cite{ATMOS} observe the deficits of detected numbers of the $\mu$-neutrino compared to theoretical numbers for them. Another indication in favor of non-zero neutrino masses is in the cosmological analysis by dark matter.\cite{SCHRAMM} 
\par 
On the other hand, terrestrial neutrino experiments searching for the neutrino masses and oscillations are under way. The FNAL, CHORUS and NOMAD \cite{MUTAU} experiments using the beam from accelerator search for $\nu_{\tau}$ appearance in a $\nu_{\mu}$, and KARMEN and LSND \cite{EMU} experiments using the accelerator beams are searching for $\nu_{\mu} \to \nu_e$ and $\bar{\nu}_{\mu} \to \bar{\nu}_e$ oscillations. The experiments using nuclear power reactor \cite{ENOTE} search for the disappearance of $\bar{\nu}_e$, in which $\bar{\nu}_e \to \bar{\nu}_X\ (X=\mu,\,\tau)$ transitions are expected. These experiments do not observe significantly large transitions. 
\par
In this paper, we examine neutrino masses and mixings which are consistent with current solar, atmospheric and terrestrial experiments in a framework where the three neutrinos have masses and mix each other. As known generally, the quantities observed in neutrino oscillation phenomena are the product of the mixing matrix for charged leptons and that for neutrinos which just corresponds to the CKM (Cabibbo-Kobayashi-Maskawa) matrix for quarks. Then, in order to study the mixing of neutrino, we have to know the mixing behavior of charged leptons. In our recent analysis,\cite{TESHIMA} we have discussed the quark mass hierarchy and mixing using the universal Yukawa coupling framework with small violations. For charged leptons, we can adopt the same mass matrices as quarks.  But, we do not assume such mass matrices for neutrinos, because mechanisms producing the neutrino mass, such as seesaw mechanism,\cite{GELLMANN} are considered for neutrinos in contrast to the quarks and charged leptons. Though there are many schemes for neutrino mass matrices which are suggested from the various models beyond the SM, we do not, in this paper, consider these schemes. After the experimental analyses for neutrino oscillation, we will discuss the neutrino mass mixing matrix and the neutrino mass matrix, using the results for neutrino oscillation obtained. 

\section{Neutrino oscillation}
Weak currents for the interactions producing and absorbing neutrinos are described as 
\begin{equation}
J_{\mu}=2\sum^3_{\alpha=1}\bar{l}^{\alpha}_L\gamma_{\mu}\nu^{\alpha}_L=2\sum^3_{a=1}\bar{l}^a_L\gamma_{\mu}U_{l_ab}\nu^b_L,
\end{equation}
where $l^{\alpha}_L$ and $\nu^{\alpha}_L$ are the weak-interaction eigenstates for charged leptons and neutrinos respectively, and  $l^a_L$ and $\nu^b_L$ are the mass eigenstates for them, which are transformed from $l^{\alpha}_L$ and $\nu^{\alpha}_L$ by the unitary matrices $U_l$ and $U_{\nu}$ as
\begin{equation}
\nu^a_L=(U_{\nu})_{a\alpha}\nu^{\alpha}_L, \ \ \ l^a_L=(U_l)_{a\alpha}l^{\alpha}_L.
\end{equation}
$U$ is the unitary matrix corresponding to the CKM matrix $V_{CKM}^{\dagger}$ for quarks defined by
\begin{equation}
U=U_lU^{\dagger}_{\nu}.
\end{equation}
In this paper, we do not consider the effects from the neutral current. It should be noted that the neutrino  emitted in the weak interaction together with the charged lepton $l_a$ is not the weak-interaction eigenstate $\nu^{\alpha}_L$ or mass eigenstate $\nu^a_L$ but the $\nu^{l_a}_L \equiv U_{l_ab}\nu^b_L$, and also the neutrino absorbed in the weak interaction together with the charged lepton $l_a$ is the same. The unitary matrices $U_l$ and $U_{\nu}$ transform the mass matrices for charged lepton mass matrix $M_l$ and neutrino one $M_{\nu}$ to diagonal mass matrices as
\begin{eqnarray}
&& U_lM^lU^{-1}_l={\rm diag}[m_e, m_{\mu}, m_{\tau}], \nonumber \\
&& U_{\nu}M^{\nu}U^{-1}_{\nu}={\rm diag}[m_1, m_2, m_3].
\end{eqnarray}
\par
For charged lepton sector, the values of $m_e$, $m_{\mu}$ and $m_{\tau}$ are well known. The unitary matrix $U_l$ is determined by the values of $m_e$, $m_{\mu}$ and $m_{\tau}$ and mass matrix for charged leptons. The charged lepton mass matrices are considered to be the same as those of quarks. There are many models for the quark mass matrices, but these are classified to two types: One type is that the quark mass matrices $M^q$ to be $M^q \simeq {\rm diag}[m^q_1, m^q_2, m^q_3]$ and thus the unitary matrices transforming them to the diagonal mass matrices $U^q$ to be $U^q \simeq 1$.\cite{LEPTON MODEL1} Another type is that the quark mass matrices are universal Yukawa coupling or democratic type ones. In our universal Yukawa coupling framework,\cite{TESHIMA} the quark mass matrix without $CP$ violation phases is described as follows: 
\begin{equation}
M^q=\Gamma^q\left(
        \begin{array}{ccc}
        1 & 1-\delta_1^q & 1-\delta_2^q \\
        1-\delta_1^q & 1 & 1-\delta_3^q \\
        1-\delta_2^q & 1-\delta_3^q & 1 
        \end{array} \right), 
 \ \ \ \ \ \ \delta_i \ll 1,        
\end{equation}
and the unitary matrices are $U^q \simeq T_0$,\cite{LEPTON MODEL2,TESHIMA} 
where $T_0$ is 
\begin{equation}
  T_0 = \left(
        \begin{array}{ccc}
        \frac1{\sqrt{2}} & -\frac1{\sqrt{2}} & 0 \\
        \frac1{\sqrt{6}} & \frac1{\sqrt{6}} & -\frac2{\sqrt{6}} \\
        \frac1{\sqrt{3}} & \frac1{\sqrt{3}} & \frac1{\sqrt{3}} 
        \end{array} \right).
\end{equation}
Thus, we adopt the next two types of $M^l$ and $U_l$ for mass matrix and mixing unitary matrix of charged leptons,
$$
M^l \sim {\rm diag}[m_e,\ m_{\mu},\ m_{\tau}], \ \ \ \ U_l \simeq 1, \quad\quad\quad \eqno{(7{\rm a})}
$$
$$ 
M^l=\Gamma^l\left(
        \begin{array}{ccc}
        1 & 1-\delta_1^l & 1-\delta_2^l \\
        1-\delta_1^l & 1 & 1-\delta_3^l \\
        1-\delta_2^l & 1-\delta_3^l & 1 
        \end{array} \right),\ \ \ (\delta_i \ll 1), \ \ \ U_l \simeq T_0. \eqno{(7{\rm b})}
$$
\par
For neutrino sector, on the other hand, the mass values are not known precisely, and the mechanism producing neutrino masses is considered to be different from that of quarks and charged leptons, such as neutrino masses are produced by the seesaw mechanism\cite{GELLMANN}. Thus, in present analysis, we put the mixing unitary matrix for neutrinos in free parameters as 
\addtocounter{equation}{1}
\begin{eqnarray}
 & & U_{\nu}=\left(
      \begin{array}{ccc}
      c_{12}^{\nu}c_{13}^{\nu} & s_{12}^{\nu}c_{13}^{\nu} & s_{13}^{\nu} \\
      -s_{12}^{\nu}c_{23}^{\nu}-c_{12}^{\nu}s_{23}^{\nu}s_{13}^{\nu} & c_{12}^{\nu}c_{23}^{\nu}-s_{12}^{\nu}s_{23}^{\nu}s_{13}^{\nu} & s_{23}^{\nu}c_{13}^{\nu} \\
      s_{12}^{\nu}s_{23}^{\nu}-c_{12}^{\nu}c_{23}^{\nu}s_{13}^{\nu} & -c_{12}^{\nu}s_{23}^{\nu}-s_{12}^{\nu}c_{23}^{\nu}s_{13}^{\nu} & c_{23}^{\nu}c_{13}^{\nu} 
      \end{array}\right), \nonumber \\
 & & \quad\quad\quad c_{ij}^{\nu}=\cos{\theta}^{\nu}_{ij},\ \  s_{ij}^{\nu}=\sin{\theta}^{\nu}_{ij},  
\end{eqnarray}  
where we neglect the {\it CP} violation phases. We will determine the mixing parameters in this matrix using the $U_l$ obtained in mass matrix for charged leptons and the $U$ obtained in the following experimental analysis. The values of neutrino masses are not known precisely, but we know these to be small as suggested by cosmological analysis by dark matter\cite{SCHRAMM} 
\begin{equation}
m_1+m_2+m_3 \sim {\rm several \ eV}.
\end{equation}

\par
In order to simplify the discussion in the following analysis, we classify the CKM matrix $U$ for leptons to next three cases:
\begin{flushleft}
\ \ (I)\ \ $U \simeq 1$,\\
\ \ (II)\ \ $U \simeq T_0,\ \  {\rm or} \ \ T_0^{\dagger}$,\\
\ \ (III)\ \ $U \simeq$ forms other than (I) and (II).  
\end{flushleft}
Case (I) corresponds to the small mixing among $\nu_e$, $\nu_{\nu}$ and $\nu_{\tau}$, and $U \simeq T_0$ case in (II) corresponds to that in which the charged lepton mass matrix is nearly universal coupling and the neutrino mass matrix is nearly diagonal, or $T_0^{\dagger}$ case does to that in which the charged lepton mass matrix is nearly diagonal and the neutrino mass matrix is nearly universal coupling.

\par
Probabilities for the transitions  $\nu_{l_a} \to \nu_{l_b}$ are given by 
\begin{eqnarray}
P(\nu_{l_a}\to\nu_{l_b})&=&|<\nu_{l_b}(t)|\nu_{l_a}(0)>|^2 = \delta_{l_a l_b}+p_{\nu_{l_a}\to\nu_{l_b}}^{12}S_{12}+p_{\nu_{l_a}\to\nu_{l_b}}^{23}S_{23}+p_{\nu_{l_a}\to\nu_{l_b}}^{31}S_{31},\nonumber \\
&&p_{\nu_{l_a}\to\nu_{l_b}}^{12}=-2\delta_{l_al_b}(1-2U_{l_a3}^2)+2(U_{l_a1}^2U_{l_b1}^2+U_{l_a2}^2U_{l_b2}^2-U_{l_a3}^2U_{l_b3}^2), \nonumber \\       
&&p_{\nu_{l_a}\to\nu_{l_b}}^{23}=-2\delta_{l_al_b}(1-2U_{l_a1}^2)+2(-U_{l_a1}^2U_{l_b1}^2+U_{l_a2}^2U_{l_b2}^2+U_{l_a3}^2U_{l_b3}^2), \nonumber \\
&&p_{\nu_{l_a}\to\nu_{l_b}}^{31}=-2\delta_{l_al_b}(1-2U_{l_a2}^2)+2(U_{l_a1}^2U_{l_b1}^2-U_{l_a2}^2U_{l_b2}^2+U_{l_a3}^2U_{l_b3}^2) ,
\end{eqnarray}
where subscripts $l_a$, $l_b$ describe $e$, $\mu$, $\tau$ and 
\begin{equation}
S_{ij}=\sin^21.27\frac{\Delta m^2_{ij}}{E}L,
\end{equation}
measuring $\Delta m^2_{ij}=m^2_i-m^2_j$, $E$ and $L$ in unit eV$^2$, MeV and m, respectively. From the unitarity of $U$, we get the relations 
\begin{equation}
p^{ij}_{\nu_{l_a}\to\nu_e}+p^{ij}_{\nu_{l_a}\to\nu_{\mu}}+p^{ij}_{\nu_{l_a}\to\nu_{\tau}}=0, \ \ \ \ i,\ j=1,2,3, \ \ \ \ l_a=e, \mu, \tau.
\end{equation}
\par
Because the expressions (10) for the probabilities $P(\nu_{l_a}\to\nu_{l_b})$ are not useful for our analysis, we rewrite them to the following special forms. If we can set $\Delta m^2_{12} \ll \Delta m^2_{23} \simeq \Delta m^2_{13}$ which is observed in the following discussion, the $P(\nu_e\to\nu_e)$'s are rewritten as follows;
\begin{eqnarray}
P(\nu_e\to\nu_e)&=&1-2(1-2U_{e3}^2-U_{e1}^4-U_{e2}^4+U_{e3}^4)S_{12}-4U_{e3}^2(1-U_{e3}^2)S_{23}, \nonumber \\
P(\nu_{\mu}\to\nu_{\mu})&=&1-2(1-2U_{\mu3}^2-U_{\mu1}^4-U_{\mu2}^4+U_{\mu3}^4)S_{12}-4U_{\mu3}^2(1-U_{\mu3}^2)S_{23}, \nonumber \\
P(\nu_\tau\to\nu_\tau)&=&1-2(1-2U_{\tau3}^2-U_{\tau1}^4-U_{\tau2}^4+U_{\tau3}^4)S_{12}-4U_{\tau3}^2(1-U_{\tau3}^2)S_{23}, \nonumber \\   
P(\nu_\mu\to\nu_e)&=&P(\nu_e\to\nu_\mu)=2(U_{\mu1}^2U_{e1}^2+U_{\mu2}^2U_{e2}^2-U_{\mu3}^2U_{e3}^2)S_{12}+4U_{e3}^2U_{\mu3}^2S_{23}, \nonumber \\
P(\nu_\tau\to\nu_e)&=&P(\nu_e\to\nu_\tau)=2(U_{\tau1}^2U_{e1}^2+U_{\tau2}^2U_{e2}^2-U_{\tau3}^2U_{e3}^2)S_{12}+4U_{e3}^2U_{\tau3}^2S_{23}, \nonumber \\
P(\nu_\tau\to\nu_\mu)&=&P(\nu_\mu\to\nu_\tau)=2(U_{\tau1}^2U_{\mu1}^2+U_{\tau2}^2U_{\mu2}^2-U_{\tau3}^2U_{\mu3}^2)S_{12}+4U_{\mu3}^2U_{\tau3}^2S_{23}.\quad 
\end{eqnarray}
If we adopt the case $U \simeq T_0$ in (II), the  $P(\nu_e\to\nu_e)$'s are rewritten on the assumption $\Delta m^2_{12} \ll \Delta m^2_{23} \simeq \Delta m^2_{13}$ as follows:
$$
\begin{array}{l}
P(\nu_e\to\nu_e)=1-S_{12}, \ \ \ P(\nu_\mu\to\nu_\mu)=1-\frac19S_{12}-\frac89S_{23}, \\
P(\nu_\tau\to\nu_\tau)=1-\frac49(S_{12}+2S_{23}),\ \ \ P(\nu_\mu\to\nu_e)=P(\nu_e\to\nu_\mu)=\frac13S_{12}, \\
P(\nu_\tau\to\nu_e)=P(\nu_e\to\nu_\tau)=\frac23S_{12}, \ \ \ P(\nu_\tau\to\nu_\mu)=P(\nu_\mu\to\nu_\tau)=-\frac29S_{12}+\frac89S_{23},
\end{array}
\eqno{(14{\rm a})}
$$
and for case $U \simeq T_0^{\dagger}$ in (II) as follows:
$$
\begin{array}{l}
P(\nu_e\to\nu_e)=1-\frac13S_{12}-\frac89S_{23}, \ \ \ P(\nu_\mu\to\nu_\mu)=1-\frac13S_{12}-\frac89S_{23}, \\
P(\nu_\tau\to\nu_\tau)=1-\frac89S_{23},\ \ \ P(\nu_\mu\to\nu_e)=P(\nu_e\to\nu_\mu)=\frac13S_{12}+\frac49S_{23}, \\
P(\nu_\tau\to\nu_e)=P(\nu_e\to\nu_\tau)=\frac49S_{23}, \ \ \ P(\nu_\tau\to\nu_\mu)=P(\nu_\mu\to\nu_\tau)=\frac49S_{23}. 
\end{array}
\eqno{(14{\rm b})}
$$
\addtocounter{equation}{1}

\section{Analyses for neutrino oscillation by experimental data}
In solar neutrino experiments, there are two results which correspond to the case considering the MSW effects \cite{MSW} and the case not considering them: 
\par\noindent
\hspace{1.5cm} the case considering MSW effect\cite{SOLAR2}
$$
\ \ \ \ \ \ \ \left\{
          \begin{array}{l}
                \Delta m^2_{21} \sim 6 \times 10^{-6} {\rm eV}^2, \\
                |p^{12}_{\nu_e\to\nu_e}| \sim 7 \times 10^{-3}, 
          \end{array} \right. \eqno{(15{\rm a})} 
$$
$$
\ \ \ \ \ \ \ \left\{
          \begin{array}{l}
                \Delta m^2_{21} \sim 8 \times 10^{-6} {\rm eV}^2, \\
                |p^{12}_{\nu_e\to\nu_e}| \sim 0.6, 
          \end{array} \right. \eqno{(15{\rm b})}           
$$
\par \noindent
\hspace{1.5cm} the case of vacuum oscillation (not considering MSW effect)\cite{SOLAR3}
$$
   \left\{ \begin{array}{l}
                \Delta m^2_{21} \sim 10^{-10} {\rm eV}^2 ,\\
                |p^{12}_{\nu_e\to\nu_e}| \ge 0.5.
          \end{array} 
   \right. \eqno{(15{\rm c})}
$$
\addtocounter{equation}{1}
From these results and the cosmological suggestion (9), we can conclude that the masses $m_1$ and $m_2$ are very small compared with $m_3$, 
\begin{equation}
m_1 \sim m_2 \ll m_3 \sim {\rm  several\ eV}.
\end{equation}
Atmospheric experiments\cite{ATMOS} detect the ratio
\begin{equation}
R \equiv \frac{R_{\rm expt}(\mu/e)}{R_{\rm MC}(\mu/e)}=\frac{P(\nu_\mu\to\nu_\mu)+\frac{N_{e}}{N_{\mu}}P(\nu_e\to\nu_\mu)+\frac{N_{\tau}}{N_{\mu}}P(\nu_\tau\to\nu_\mu)}{P(\nu_e\to\nu_e)+\frac{N_{\mu}}{N_e}P(\nu_\mu\to\nu_e)+\frac{N_{\tau}}{N_e}P(\nu_\tau\to\nu_e)}\sim 0.6,
\end{equation}
where $N_{l_a}$ is the flux of the original $\nu_{l_a}$'s  which are produced at the production point somewhere in the atmosphere.  $N_{\mu}/N_e \sim 2$ is obtained from Monte Carlo calculation. Because $\tau$ lepton production is very small at the energy range of these atmospheric experiments, $N_{\tau}/N_e$ and $N_{\tau}/N_{\mu}$ are considered to be very small. Thus we get the result   
\begin{equation}
\frac{P(\nu_\mu\to\nu_\mu)+\frac{1}{2}P(\nu_e\to\nu_\mu)}{P(\nu_e\to\nu_e)+2P(\nu_\mu\to\nu_e)}\sim 0.6.
\end{equation}
The results for terrestrial experiments are as follows: 
\par\noindent
\hspace{1.5cm} accelerator experiment for $\nu_\mu \to \nu_\tau$,\cite{MUTAU}
\begin{equation}
  P(\nu_\mu\to\nu_\tau) < 4\times10^{-3},
\end{equation}  
\hspace{1.5cm} accelerator experiment for $\nu_\mu \to \nu_e$ and $\bar{\nu}_\mu \to \bar{\nu}_e,$\cite{EMU}  
\begin{equation}  
P(\nu_\mu\to\nu_e) < 4.8\times10^{-2},\ \ \ \ \ P(\bar{\nu}_\mu\to\bar{\nu}_e) < 6.2\times10^{-3},
\end{equation}
\hspace{1.5cm} reactor experiments for $\bar{\nu}_e \not\to \bar{\nu}_e$,\cite{ENOTE} 
\begin{equation}
1-P(\bar{\nu}_e\to\bar{\nu}_e) < 2\times10^{-2}. 
\end{equation}

\par
Now, we search the CKM matrix $U$ of leptons consistent with above experimental results. Though the neutrino masses are suggested like Eq.~(9) by cosmological analysis, we consider two possibilities of values of neutrino masses. One is that $m_3 < 0.1\times{\rm several \ eV}$ but not so small (not $m_3\ll 0.1\times{\rm several \ eV}$) and the other is that  $m_3 \sim {\rm several \ eV}$.
\par\noindent
(A).\ \  $m_3 < 0.1\times{\rm several \ eV}$ case
\par\noindent
(A-1).\ \ Terrestrial experiments 
\par
$S_{12}$, $S_{13}$ and $S_{23}$ are very small ($\ll1$) for terrestrial experiments ($1.27\Delta m_{ij}^2 L/E \ll \pi/2$), thus these experiments are satisfied independently of the mixing matrix $U$ for leptons.\\
(A-2).\ \  Solar neutrino experiment
\par
In solar neutrino experiments, if we take the case considering MSW effect, there are two cases (15a) and (15b). The former is called as a small-mixing solution and the latter a large-mixing solution.\\
(A-2-a).\ \  (15a) case
\par
In this case $|p^{12}_{\nu_e\to\nu_e}|\sim7\times10^{-3}$, then from Eq.~(13), one can get
\begin{equation}2(1-2U^2_{e3}-U^4_{e1}-U^4_{e2}+U^4_{e3})\sim7\times10^{-3}.
\end{equation}
As found easily, this result is satisfied by case (I), but not satisfied by case (II). \\
(A-2-b).\ \ (15b) case
\par
In this case, 
\begin{equation}
2(1-2U^2_{e3}-U^4_{e1}-U^4_{e2}+U^4_{e3})\sim0.6.
\end{equation}
This result is satisfied by case (II), but not satisfied by case (I). \\
(A-2-c).\ \ (15c) case
\par
In the case of vacuum oscillation (15c) not considering the MSW effect,  expression of $P(\nu_e\to\nu_e)$ in Eq.~(13) or Eq.~(14) is used for analysis. In these experiments, $\Delta m^2_{12}\sim10^{-10}{\rm eV^2}$ then $S_{12}\sim1$, and $|p^{12}_{\nu_e\to\nu_e}| \ge 0.5$. In this case, the case (I) is ruled out but the case (II) is not ruled out. \\
(A-3).\ \ Atmospheric experiments.
\par
Next, we consider the atmospheric neutrino experiments Eq.~(18). If we adopt the solutions (15a) or (15b), $S_{12}$ in $P(\nu_\mu\to\nu_\mu)$'s is about $\sim 0.01$, because $E \sim 1{\rm GeV}$ and $L \sim 10^4{\rm km}$ for atmospheric neutrinos. $S_{23}$ is averaged as $\frac12$, because $\dis{1.27\frac{\Delta m^2_{23}}{E}L}$ is larger than $\pi/2$ in our assumption for the value of $m_3$ to be not $\ll 0.1 \times$ several eV. Then 
$$
\frac{P(\nu_\mu\to\nu_\mu)+\frac12P(\nu_e\to\nu_\mu)}{P(\nu_e\to\nu_e)+2P(\nu_\mu\to\nu_e)} \sim 0.6, \eqno{(18)}
$$
\begin{eqnarray}
P(\nu_\mu\to\nu_\mu) & \sim & 1+p_{\nu_\mu\to\nu_\mu}^{12}S_{12}+\frac12(p_{\nu_\mu\to\nu_\mu}^{23}+p_{\nu_\mu\to\nu_\mu}^{31}), \nonumber \\
P(\nu_e\to\nu_e) & \sim & 1+p_{\nu_e\to\nu_e}^{12}S_{12}+\frac12(p_{\nu_e\to\nu_e}^{23}+p_{\nu_e\to\nu_e}^{31}), \nonumber\\
P(\nu_\mu\to\nu_e) & = & P(\nu_e\to\nu_\mu) \sim p_{\nu_\mu\to\nu_e}^{12}S_{12}+\frac12(p_{\nu_\mu\to\nu_e}^{23}+p_{\nu_\mu\to\nu_e}^{31}).
\end{eqnarray}
As shown in numerical analysis, there are solutions for $U_{e3}$'s satisfying these constraints. Here it should be noted that the case (I) does  not satisfy these constraints. If we adopt the case (15c), the $S_{12}$ terms are negligible. Then we get such relation as
\begin{equation}
\{U^2_{\mu3}(1-U^2_{\mu3})-0.6U^2_{e3}(1-U^2_{e3})+0.7U^2_{e3}U^2_{\mu3}\}S_{23}\sim0.1. 
\end{equation}
This result is not satisfied by case (I) but is satisfied by case (II), because we assume that $m_3$ is small but not so small i.e. not $\ll0.1\times\ {\rm several\ eV}$.  
\par 
Thus, if $m_3 < 0.1\times{\rm several \ eV}$ but not so small (not $m_3\ll 0.1\times{\rm several\ eV}$), there are solutions satisfying the terrestrial, solar (large-mixing solution) and atmospheric neutrino experiments in cases (II) and (III), but not in case (I). The possibility $U \simeq T_0$ in (II) has been analysed by the literature in Ref.~\cite{FRITZSCH}. 
\par \noindent 
(B).\ \  $m_3 \sim {\rm several \ eV}$ case \\
(B-1).\ \ Terrestrial experiments  
\par
Next, we consider the values of neutrino mass in the case Eq.~(9). In this case, we must consider the terms proportional to $S_{23}$ in Eq.~(13) for terrestrial experiments, because the term $S_{23}$ is not so small. We get the restrictions for $U_{l_ai}$'s,  
\begin{equation}
U^2_{e3}(1-U^2_{e3})<10^{-2},\ \ \ U^2_{e3}U^2_{\mu3}<10^{-2}, \ \ \ U^2_{\mu3}U^2_{\tau3}<10^{-3}.
\end{equation} 
There are 3 possibilities satisfying these restrictions,
$$
U_{e3}\sim0,\ \ U_{\mu3}\sim0,\ \ U_{\tau3}\sim1,\ \ \eqno{(27{\rm a})} 
$$
$$  
U_{e3}\sim0,\ \ U_{\mu3}\sim1,\ \ U_{\tau3}\sim0,\ \ \eqno{(27{\rm b})} 
$$
$$
U_{e3}\sim1,\ \ U_{\mu3}\sim0,\ \ U_{\tau3}\sim0,\ \ \eqno{(27{\rm c})}
$$
\addtocounter{equation}{1}
where the unitarity relation $U^2_{e3}+U^2_{\mu3}+U^2_{\tau3}=1$ is used. In this case, we notice that the case (II) is ruled out. \\
(B-2-a). \ \ (15a) case
\par
Next we consider the solar neutrino experiments. Because 3 possibilities (27a-c) make the third term of $P(\nu_e\to\nu_e)$ in Eq.~(13) very small, as the case (A),  Eq.(22) must be satisfied. This restriction is satisfied by the possibilities (27a), (27b) and (27c) because of the unitarity $U^2_{e1}+U^2_{e2}+U^2_{e3}=1$.\\     
(B-2-b,c). \ \ (15b) and (15c) case
\par
In these large-mixing solution cases,  Eq.(23) must be satisfied. These are not satisfied by the possibility (27c). Even in possibility (27a), it should be noticed that the case (I) does not satisfy this restriction.\\ 
(B-3).\ \ Atmospheric experiments
\par
In this case, the circumstance is similar to (A-3) case. Then the constraints (18) and (24) must be satisfied. If $\Delta m^2_{12} \sim 10^{-5}{\rm eV}$ in the solution for solar neutrino experiment is used, then $S_{12} \sim 0.01$ and a solution satisfying above constraints is not obtained. But, if we adopt the value for $\Delta m^2_{12}$ to be $\sim 1$, we can take a solution. The solution for $U_{e3}$'s satisfying these constraints are shown later. Here it should be noted that the case (I) does  not satisfy these constraints. Because, for the (15c) case, we get the relation,
$$
\{U^2_{\mu3}(1-U^2_{\mu3})-0.6U^2_{e3}(1-U^2_{e3})+0.7U^2_{e3}U^2_{\mu3}\}S_{23}\sim0.1, \eqno{(25}
$$
then this result is not satisfied by case (I).
\par
Thus, if $m_3 \sim {\rm several \ eV}$, the case (II) is ruled out from the terrestrial experiments. Furthermore, the case (I) is also ruled out from the atmospheric experiments. Then in this $m_3 \sim {\rm several \ eV}$ case, both (I) and (II) case are ruled out. In case (III) other than (I) and (II), three possibilities (27a), (27b) and (27c) must be satisfied. If we adopt the large-mixing solution (15b) and (15c) for solar neutrino experimental results, the possibility (27c) is ruled out. The possibility (27b) has been discussed by a literature in \cite{BILENKY}.
\par
We next search the allowed values of $U_{ei}$'s  satisfying the constraints (26), (18) and (24), which correspond to the large-mixing solution in solar neutrino experiment (15b) and (15c) cases. Though the values of $\Delta m^2_{12}$ in the large-mixing solution in solar neutrino experiment are $\sim10^{-5}{\rm eV^2}$, we adopt the values for $\Delta m^2_{12}$ in Eq.~(24) to be $\sim5\times10^{-5}{\rm eV^2}$ and then $S_{12}$ to be $\sim0.3$ in order to see a mixing effect in atmospheric neutrino experiments. In recent analysis for solar neutrino experiments within matter enhanced three neutrino oscillations\cite{FOGLI}, values of the $\Delta m^2_{12}$ to be $\sim10^{-4}{\rm eV^2}$ is allowed. The typical solutions obtained are as follows:
$$
U=\left(\begin{array}{ccc}
   0.868 & 0.481 & 0.122 \\
   -0.496 & 0.855 & 0.155 \\
   -0.029 & -0.195 & 0.980
   \end{array}\right), \quad\quad\quad\quad\quad\quad\quad\quad\eqno{(28{\rm a})} 
$$
$$   
U=\left(\begin{array}{ccc}
    0.481 & 0.868 & -0.122 \\
   -0.855 & 0.496 & 0.155 \\
    0.195 & 0.029 & 0.980
   \end{array}\right), \quad\quad\quad\quad\quad\quad\quad\quad\eqno{(28{\rm b})} 
$$
\begin{eqnarray}
&&\left\{\begin{array}{l}
P(\nu_\mu\to\nu_\tau)\sim|p^{23}_{\nu_\mu\to\nu_\tau}+p^{31}_{\nu_\mu\to\nu_\tau}|= 0.093,\\ 
P(\nu_\mu\to\nu_e)\sim|p^{23}_{\nu_\mu\to\nu_e}+p^{31}_{\nu_\mu\to\nu_e}|= 0.0014,\\
1-P(\bar{\nu_e}\to\bar{\nu_e})\sim|p^{23}_{\nu_e\to\nu_e}+p^{31}_{\nu_e\to\nu_e}|= 0.059  ,\end{array} \right. {\quad\rm for\ terrestrial\ experiments} \nonumber \\
&&|p^{12}_{\nu_e\to\nu_e}|=0.70,{\quad\rm for\ solar\ experiments} \nonumber \\ 
&&R= \frac{P(\nu_\mu\to\nu_\mu)+\frac12P(\nu_e\to\nu_\mu)}{P(\nu_e\to\nu_e)+2P(\nu_\mu\to\nu_e)}=0.71.{\quad\rm for\ atmospheric\ experiment} \nonumber
\end{eqnarray} 
\addtocounter{equation}{1}
For the constraints (24), (18) and (22) in $m_3\sim {\rm several\ eV}$, which corresponds to the small-mixing solution in solar neutrino experiment (15a) case, the typical solution obtained is as follows: 
\begin{eqnarray}
U&=&
   \left(\begin{array}{ccc}
    0.113 & 0.109 & 0.988 \\
   -0.905 & -0.399 & 0.147 \\
    0.410 & -0.911 & 0.054
   \end{array}\right),\\
&&\left\{\begin{array}{l}
P(\nu_\mu\to\nu_\tau)\sim|p^{23}_{\nu_\mu\to\nu_\tau}+p^{31}_{\nu_\mu\to\nu_\tau}|= 0.0002,\\ 
P(\nu_\mu\to\nu_e)\sim|p^{23}_{\nu_\mu\to\nu_e}+p^{31}_{\nu_\mu\to\nu_e}|= 0.084,\\
1-P(\bar{\nu_e}\to\bar{\nu_e})\sim|p^{23}_{\nu_e\to\nu_e}+p^{31}_{\nu_e\to\nu_e}|= 0.095  ,\end{array} \right. {\quad\rm for\ terrestrial\ experiments} \nonumber\\
&&|p^{12}_{\nu_e\to\nu_e}|=0.0006,{\quad\rm for\ solar\ experiments}\nonumber\\ 
&&R= \frac{P(\nu_\mu\to\nu_\mu)+\frac12P(\nu_e\to\nu_\mu)}{P(\nu_e\to\nu_e)+2P(\nu_\mu\to\nu_e)}=0.80.{\quad\rm for\ atmospheric\ experiment}\nonumber 
\end{eqnarray}   

\section{Discussions}
As shown in above numerical analyses, we get the solutions (28a,b) and (29) for $U$ satisfying the present solar, atmospheric and terrestrial experiments in $m_3\sim {\rm several\ eV}$. In $m_3 < 0.1\times{\rm several\ eV}$, there are many solutions satisfying these experiments. In these solutions, case (II) is included.
\par
First, we comment on the case $m_3 < 0.1\times{\rm several\ eV}$. In this case, there are many solutions but the possibility (I) which corrsponds to the case of small mixing among $\nu_e$, $\nu_\mu$ and $\nu_\tau$ is ruled out. In these solutions, the possibility of case $U \sim T_0$ in (II) is very interesting, because the large mixing of $U$ is introduced naturally from the mixing $T_0$ of universal Yukawa coupling charged leptons and no mixing of neutrino.\cite{FRITZSCH} The possibility of case $U \sim T_0^{\dagger}$ in (II) corresponds to the mixing $T_0$ of universal Yukawa coupling neutrino and no mixing of charged lepton.  Though there are many other models producing large mixing of $U$ through the seesaw mechanism, these models use the Majolana mass matrix producing large mixing. 
\par
Next we comment on the case $m_3 \sim {\rm several\ eV}$. In this case, the case $U \sim 1$ of (I) and  cases $U \sim T_0$ and $U \sim T_0^{\dagger}$ in (II) are ruled  out. We get the typical solutions (284a, b) for the large-mixing solutions (15b) and (15c) in solar neutrino experiment. Matrices for these solutions are characterized as  
$$
U \simeq \left(\begin{array}{ccc}
   \frac{\sqrt{3}}{2} & \frac12 & 0 \\
   -\frac12 & \frac{\sqrt{3}}{2} & 0 \\
   0 & 0 & 1
   \end{array}\right),  \eqno{(30{\rm a})} 
$$
$$   
U \simeq \left(\begin{array}{ccc}
   \frac12 & \frac{\sqrt{3}}{2} & 0 \\
   -\frac{\sqrt{3}}{2} & \frac12 & 0 \\
   0 & 0 & 1
   \end{array}\right). \eqno{(30{\rm b})}   
$$
\addtocounter{equation}{1} 
The matrix of the solution (29) for the small-mixing solution (15a) in solar neutrino experiment is characterized as 
\begin{equation}
U \simeq \left(\begin{array}{ccc}
   0 & 0 & 1\\
   -\frac{\sqrt{3}}{2} & \frac12 & 0 \\
   -\frac12 & -\frac{\sqrt{3}}{2} & 0 \\
   \end{array}\right).
\end{equation}
\par
The pattern of solutions (30a,b) is explained naturally by the universal Yukawa coupling mass matrix for charged lepton and neutrino. In fact, for the solution (30a) and $U_{l} \simeq T_0$, the neutrino transformation unitary matrix is given as 
\begin{equation}
U_{\nu} \simeq T'_0 = \left(
        \begin{array}{ccc}
        \frac1{\sqrt{6}} & -\frac2{\sqrt{6}} & \frac1{\sqrt{6}} \\
        \frac1{\sqrt{2}} & 0 & -\frac1{\sqrt{2}} \\
        \frac1{\sqrt{3}} & \frac1{\sqrt{3}} & \frac1{\sqrt{3}}   
        \end{array} \right).
\end{equation}
Though this transformation matrix $T'_0$ is different from $T_0$ in the first and second rows which describe the eigenstates for small eigenvalues $m_1$ and $m_2$, but it describe the transformation matrix of universal Yukawa coupling mass matrix. 
\par
We show the mixing matrix (8) and mass matrix $M^{\nu}$ for neutrino for the solution (28a) using the charged lepton mass matrix (7b). Here we use the mass ratios for charged leptons, $\dis{\frac{m_e}{m_{\mu}}=0.004836,\ \ \frac{m_{\mu}}{m_{\tau}}=0.05946\pm0.00001}$,\cite{PDG96}\,\ then the mass matrix $M^l$ and transformation matrix $U_l$ for the charged leptons are given as follows:
\begin{eqnarray}
&& M^l=\Gamma^l\left(
        \begin{array}{ccc}
        1 & 1-0.002 & 1-0.137 \\
        1-0.002 & 1 & 1-0.113 \\
        1-0.137 & 1-0.113 & 1 
        \end{array} \right), \nonumber\\
&& U_l=\left(
      \begin{array}{ccc}
      0.673 & -0.736 & 0.073 \\
      0.455 & 0.334  & -0.826 \\
      0.584 & 0.588 & 0.559 
      \end{array}\right).         
\end{eqnarray}
Then the solution of $U_{\nu}$ and $M^{\nu}$ for (28a) is obtained as follows:
\begin{eqnarray}
&& U_{\nu}=\left(
      \begin{array}{ccc}
      0.341 & -0.822 & 0.456 \\
      0.598 & -0.184  & -0.780 \\
      0.725 & 0.539 & 0.429 
      \end{array}\right),\ \ 
      \begin{array}{ll}
      \theta_{12}=-64.5^{\circ}, & \theta_{23}=-40.1^{\circ},\\
      \theta_{13}=29.3^{\circ}, & 
      \end{array}
      \ \ \nonumber\\
&& M^{\nu}=\Gamma^{\nu}\left(
        \begin{array}{ccc}
        0.536 & 0.387 & 0.297 \\
        0.387 & 0.292 & 0.235 \\
        0.297 & 0.235 & 0.203
        \end{array} \right),\\
&&\ \quad\quad\quad {\rm for\ } \frac{m_e}{m_{\mu}}=0.033,\ \ \frac{m_{\mu}}{m_{\tau}}=0.03. \nonumber 
\end{eqnarray}
The solution of $U_{\nu}$ and $M^{\nu}$ for (28b) is obtained as follows:
\begin{eqnarray}
&& U_{\nu}=\left(
      \begin{array}{ccc}
      0.049 & -0.525 & 0.850 \\
      0.826 & -0.456  & -0.330 \\
      0.561 & 0.718 & 0.411 
      \end{array}\right),\ \ 
      \begin{array}{ll}
      \theta_{12}=-84.1^{\circ}, & \theta_{23}=-38.3^{\circ},\\
      \theta_{13}=58.6^{\circ}, & 
      \end{array}
      \ \ \nonumber\\
&& M^{\nu}=\Gamma^{\nu}\left(
        \begin{array}{ccc}
        0.335 & 0.392 & 0.223 \\
        0.392 & 0.523 & 0.300 \\
        0.223 & 0.300 & 0.173
        \end{array} \right),\\
&&\ \quad\quad\quad {\rm for\ } \frac{m_e}{m_{\mu}}=0.033,\ \ \frac{m_{\mu}}{m_{\tau}}=0.03. \nonumber 
\end{eqnarray}
For (29), $U$ is characterized as (31), then the neutrino translation unitary matrix is given as  
\begin{equation}
U_{\nu} \simeq T''_0 = \left(
        \begin{array}{ccc}
        -\frac{\sqrt{3}+\sqrt{2}}{2\sqrt{6}} & -\frac{\sqrt{3}+\sqrt{2}}{2\sqrt{6}} &\frac{2\sqrt{3}-\sqrt{2}}{2\sqrt{6}} \\
        \frac{1-\sqrt{6}}{2\sqrt{6}}  & \frac{1-\sqrt{6}}{2\sqrt{6}} & -\frac{2+\sqrt{6}}{2\sqrt{6}}\\
        \frac1{\sqrt{2}}  & -\frac1{\sqrt{2}}  & 0 
        \end{array} \right)
\end{equation} 
Thus, in this case, we cannot get the universal Yukawa coupling type solution for neutrino mass .
\par
We discussed the neutrino oscillation which are consistent with current solar, atmospheric and terrestrial experiments in a framework with three neutrinos. In the neutrino mass $m_{3}$ to be $< 0.1\times{\rm several\ eV}$, there are many solutions consistent with these experiments, but the possibility for small mixing of $U$ is ruled out. In these solutions there is the possibility of case $U \sim T_0$, which is very interesting because the large mixing of $U$ is introduced naturally from the mixing $T_0$ of universal Yukawa coupling charged leptons and no mixing of neutrino.  In the neutrino mass $m_{3}$ to be $\sim {\rm several\ eV}$, the neutrino mixing matrix $U$ must produce the large mixing between $\nu_e$-$\nu_{\mu}$ and $\nu_{\mu}$-$\nu_{\tau}$ as 
   $\dis{U \simeq \left(\begin{array}{ccc}
   \frac{\sqrt{3}}{2} & \frac12 & 0 \\
   -\frac12 & \frac{\sqrt{3}}{2} & 0 \\
   0 & 0 & 1
   \end{array}\right)}$ 
   , 
   $\dis{U \simeq \left(\begin{array}{ccc}
   \frac12 & \frac{\sqrt{3}}{2} & 0 \\
   -\frac{\sqrt{3}}{2} & \frac12 & 0 \\
   0 & 0 & 1
   \end{array}\right)}$ 
   and 
   $\dis{U \simeq \left(\begin{array}{ccc}
   0 & 0 & 1\\
   -\frac{\sqrt{3}}{2} & \frac12 & 0 \\
   -\frac12 & -\frac{\sqrt{3}}{2} & 0 \\
   \end{array}\right)}$.
The large mixing between $\nu_e$-$\nu_{\mu}$ are very interesting, because these transformation matrices are obtained naturally from the universal Yukawa coupling mass matrices for charged leptons and neutrinos. 


\end{document}